\documentclass[a4paper,12pt]{article}
\usepackage{epsfig}
\newcommand{\etal}{{\it et al.}}
\begin{document}

\title{
Single particle strength restoration and nuclear transparency 
in high $Q^2$ exclusive $(e,e^{\prime}p)$ reactions}
\author{
L. Frankfurt\\
\it School of Physics and Astronomy, Raymond and Beverly Sackler\\
\it Faculty of Exact Science, Tel Aviv University, Ramat Aviv 69978,\\
\it Tel Aviv , Israel\\
M. Strikman\\
\it Pennsylvania State University, University Park, Pennsylvania 16802\\
M. Zhalov\\
\it Petersburg Nuclear Physics Institute, Gatchina 188350, Russia} 
\date{}
\maketitle

\centerline {\bf ABSTRACT}

Quasifree $(e,e^{\prime}p)$ reactions at $Q^2\sim 0.1 GeV^2$
observed 
 a strong quenching of the single-particle valence strength. 
This  precluded unambiguous measurement of nuclear 
transparency in quasifree (e,e'p) reactions at $Q^2\geq 1\,GeV^2$.
We argue that  the high-energy nuclear transparency in the transverse kinematics
weakly depends on the probability of the short-range correlations and 
can be accurately determined using information on the cross section of the $(e,e')$ 
reaction at $x\sim 1$ and $ Q^2\sim 1\div 2 GeV^2$. 
We find that the Glauber approximation well describes E91-013 and 
NE18 (e,e'p) data at $2\leq Q^2\leq 4 GeV^2$ without any significant quenching. 
This gives  further support to our  observation 
that the quenching
of nuclear levels strongly depends on the resolution ($Q^2$) and practically disappears
at  $Q^2\geq 1\,GeV^2$.

\section{Introduction}

Quasi-free knockout (e,e'p) reactions were used for a long 
time to study nuclear structure at the energy  transfer $q_0 \leq 500 MeV$, 
for a review see \cite{mogad,SWH,Kelly:1996hd}. One of the important findings of these
studies was an observation of  suppression of the single-particle 
valence strength  as compared to the calculations using the nuclear shell model wave 
functions. This phenomenon of quenching in the low energy physics is naturally explained in the
nuclear quasiparticle theory\cite{Migdal} as a result of the single
particle strength fragmentation over wide excitation energy range due to the 
long and short range nucleon-nucleon correlations(for the recent review see  e.g. \cite{Dickhoff:1992um}). 
Recently a thorough comparison of all recent low energy  data for $^{12}C(e,e'p)$
was performed in \cite{lsfsz}. It is found that the data from different experiments 
are consistent with each other and require a very substantial reduction of the s- and p- shell
strength in $^{12}C(e,e'p)$ at $Q^2\le \,0.3\,GeV^2$ by the factor
\begin{equation}
  \eta(^{12}C)= .57\pm .02  . 
\label{lowenq}
\end{equation} 
If the quenching did not depend on $Q^2$, this finding 
would strongly affect the interpretation of the recent
(e,e'p) experiments at 
high energies and momentum transfers 
\cite{ne18,tjnaf}.
These experiments were performed
to study the nuclear transparency  ${\bf T}$
 as a function of 
the momentum transfer at $1\,GeV^2\leq Q^2\leq 7\,GeV^2$  on
several nuclei with the main goal 
%The main goal of these experiments was 
to search for the color transparency effects \cite{CT,Brodsky,Frankfurt:1994hf}.
According to the theoretical predictions\cite{frank1,frank2} the color transparency  could be expected
at these momentum transfers as a 10\% increase of the nuclear transparency with 
increase of $Q^2$.
Hence, a high precision determination of all the nuclear characteristics influencing 
evaluation of ${\bf T}$ is very important.

Experimentally, the nuclear transparency ${\bf T}$ is defined as  the ratio 
of the observed cross section to the cross section calculated in the 
plane-wave impulse approximation(PWIA).
The delicate point here is to estimate precisely  how large the latter is 
for the kinematics of the particular experiment. The PWIA cross section is 
\begin{equation}
\sigma_{pwia}=F_{kin}\sigma^{ep}_{cc1}
\int \ S({\bf k},E)\,d^3{ k}\,d\,E,
\label{sigpw}
\end{equation}
where $F_{kin}$ is the kinematic factor  
and
 $\sigma^{ep}_{cc1}$ \cite{deforest}  is the off-shell extrapolation of the
elastic $ep$ cross section.  
  If the momenta  of bound nucleons, ${\bf k}$, and excitation energies  of the residual nuclei,
$E$, in (e,e'p)  are not too large,
there are practically no differences between different models 
for off-shell extrapolation of 
 $\sigma^{ep}$. Hence, 
the main problem is to determine accurately enough the integral over  
the  nuclear spectral function $S({\bf k},E)$.
In high $Q^2$  (e,e'p) experiments \cite{ne18,tjnaf}
the  $S({\bf k},E)$ was calculated in the independent particle shell model.
To correct this calculation for  the missing single particle strength 
in the kinematics of the experiment 
an additional correction factor $f(A)$ was introduced, leading to 
\begin{equation}
\int 
S({\bf k},E)\,d\,E\,d^3\,{ k}=f(A)\cdot
\int 
S_{IPSM}({\bf k},E)d\,E\,d^3\,{ k}.
\label{corcor}
\end{equation}
In particular, the values
\begin{equation}
f(^{12}C)=0.9,\,\, \, \, f(^{56}Fe)=0.82, \,\,  \,\, f(^{197}Au)=0.78,
\label{corfac}
\end{equation}
 were
used in \cite{ne18,tjnaf} for extracting the nuclear transparency from the data.

It was recently pointed out in \cite{lsfsz} that   interpreting
 transparency measurements at high energies performed
in the   transverse kinematics, with the cuts on the momentum of the struck 
nucleon and the energy of the produced system, 
requires a re-evaluation 
of the quenching which was observed in the low energy domain.
In particular,  if one would use the same quenching for excitation of
$s-,p-$ hole states in the carbon as the one observed at the low 
$Q^2\le 0.3\, GeV^2$  (eq.\ref{lowenq}), the transparency ${\bf T}$ 
for the $^{12}C(e,e'p)$ s- and p-valence state region is about 0.8 for
$Q^2 = 1\, GeV^2$.
 This number is much higher than the predictions of Glauber theory which 
should be a reasonable approximation for 
$Q^2\ge 1\, GeV^2$ and should be a  
very good approximation for $E_N\geq 1$ GeV, corresponding to 
$Q^2\geq 2\, GeV^2$. At the same time we demonstrated that 
the carbon data at $Q^2=1\, GeV^2$ for the differential $(e,e'p)$ cross section
appeared to be consistent with the Glauber calculation provided one assumes
a strong reduction of the quenching effect at large $Q^2$.
In particular we used the NE-18 differential (e,e'p) cross sections for carbon  to determine
the quenching factor for $Q^2=1\, GeV^2$ to be $\sim 0.9$.

We further argued that
 a $Q^2$ dependence of 
quenching should be a natural phenomenon
reflecting  transition from low $Q^2$ interactions where
photon  interacts with quasiparticles to the interaction with  nucleons
at larger $Q^2\geq 1\, GeV^2$. However, we see  no reasons for a noticeable A-dependence
of the correlation correction at high $Q^2$ and $A\ge 12$.
Indeed the main source of the $A$ dependence at high $Q^2$  would be the
$A$-dependence of the short-range correlation contribution. 
According to the  the analysis of
high $Q^2$, $x>1$ data(see e.g. \cite{fsscor}) the effect of the short range correlations changes by $\le 20\%$
between A=12 and A=208. However,  as we show below,this contribution itself
in the considered integral is just a few \%($\sim (5\div 7)\%$ for carbon).

In this paper we extend the analysis of \cite{lsfsz} in several directions.
We analyze the
transparency measured recently in \cite{tjnaf} for a range of nuclei, 
focusing at 
$Q^2=1.8\, GeV^2$ for which both integrated cross sections and differential 
cross sections are available. Our  choice of $Q^2$ is motivated  by 
a very good understanding of $NN$ interactions
 for the corresponding energy of the ejected nucleon $E_p\approx 1 GeV$ -
the Glauber theory is known to describe numerous data on 
elastic and quasielastic $pA$ interactions 
at this energy with a typical accuracy of few percents, see review in 
\cite{vorobiov}. Also, due to a weak energy 
dependence of $\sigma_{pN}$ between 
$E_p\approx 1\, GeV$ and $E_p\approx 2\, GeV$,
and the smallness of the color transparency effects for  
the $Q^2\leq 4\, GeV^2$ range, 
one expects a very weak dependence of transparency on $Q^2$ for 
$2\,GeV^2 \leq Q^2 \leq 4 \,GeV^2$. This is certainly consistent with the data. Hence
adding higher $Q^2$ data would not add much to
the main thrust of  our analysis. 
 We observe that measurements of ${\bf T}$ in the transverse kinematics of $x=1$
are not sensitive to the high momentum component of the 
nuclear wave function since the cross section is proportional to
$\int  {S({\bf k},E)\over k} {d^3k} dE$ rather than to $\int S({\bf k},E) d^3k dE$.
%, where $ S_A(k,E)$ is the nuclear spectral function. 
Further reduction in 
the uncertainties is reached by using information on the cross sections
of $(e,e')$ scattering at $ x=1$ and $Q^2\sim 1\, GeV^2$  measured
at Jlab \cite{arrington} which allows to determine independently 
$\int  {S({\bf k},E)\over k} {d^3k } dE$ with an  accuracy of few \%.
Using this information we calculated the transparency for the kinematics
of the  E91-013 experiment for carbon, iron and gold and find that with an 
appropriate normalization of the impulse approximation  cross section we obtain
a very good description of the data.

We  also check our conclusions about 
the noted reduction of quenching by comparing
the results of our calculations with the $(e,e'p)$ data from the 
Jlab experiment \cite{dutta} for the differential cross sections.
We will show that excellent agreement is observed,
 without any adjusted parameters, 
for the region $k_N\leq 200\, MeV/c$,  where contribution of 
the short-range correlations
is small. This provides  a very strong new evidence for the practical
 disappearance of the   quenching at large $Q^2$.

 In the end of the paper we consider 
 implications for optimizing searches for 
color transparency  in high $Q^2$  $A(e,e'p)$ processes.
Numerical predictions for kinematics where the  onset of 
the color transparency is expected will be presented elsewhere.

\section{Definition of transparency - how large is the impulse approximation}
\label{norma}

Current experiments which study
 nuclear transparency perform measurements in a restricted region of recoil
nuclear momenta and excitation energies.
Hence to convert the measured cross section to the value of transparency
${\bf T}$ 
it is necessary to
consider the ratio:
\begin{equation}
{\bf T}=\frac {1} {\sigma_{pwia}}
{\int_{\Delta^3 k}\,d^3\,{ k}\\
\int_{\Delta E}\, d\,E\ d\,\sigma_{exp}({\bf k},E)}
\equiv
{\int_{\Delta^3 k}\,d^3\,{ k}\\
\int_{\Delta E}\, d\,E\,
{S^{exp}({\bf k},E)}\\
\over\\
{
\int_{\Delta^3 k}\,d^3{ k}\\
\int_{\Delta E}\, d\,E\,
S({\bf k},E)
}}.
\label{tne18}
\end{equation} 
The quantities  $\Delta^3 k$  and $\Delta E$ in (\ref{tne18})  define
the  ranges in  missing
momentum ${\bf k}={\bf q}-{\bf p}$  and missing energy  $E=\nu-T_p$.
The value of transparency ${\bf T }$ is known to depend appreciably on 
the  excitation energy, the missing momentum and angle between 
${\bf k}$ and ${\bf q}$\cite{Golubeva:1998at,frank1}.

In the kinematics of the NE18 and E91-013 experiments  $ |k|$  and $\Delta E$ 
 were restricted
by  300 MeV/c and 80 MeV. Besides,
the transverse kinematics of the experiments corresponded approximately to
 $k_3=0$. 
Account for the kinematics of the quasielastic process leads
to the following relationship  between $k_{3}$ and the 
Bjorken scaling variable 
$ x=\frac{Q^2}{2m_N\nu}$  valid at sufficiently large $Q^2$   
\begin{eqnarray}
\frac{k_{3}}{m_{N}}=
\frac {1-x+(M_{A}-M^*_{A-1}-m_N)/m_{N}}{\sqrt {(1+4m_{N}^2/Q^2)}}.
\label{eq:k_{3}}
\end{eqnarray}
Thus, the    $k_3 \approx 0$ condition implies in the kinematics of 
the NE18 and E91-013 experiments $x\approx 1$,   and that
the main contribution to the cross section is given by the region
\begin{equation}
\left({\vec{k}\cdot \vec{q}\over |\vec{q}|}\right)^2\ll k^2.
\label{trans}
\end{equation}
Obviously, if no restrictions other than $k_3\approx 0$ were imposed
we would obtain the quasielastic contribution to the total cross
section of the $(e,e')$ cross section at $x\approx 1$ for the same $Q^2$.
At sufficiently high $Q^2$ this cross section
is proportional to 
$$S(k_3=0)\equiv \int S({\bf k_t},k_3=0,E)d^2{ k_t} dE$$
 which coincides with the
integrated spectral function $F(y)$ in the y-scaling models for
$y=0$. 
An important feature of this integral
 is that it has a much smaller contribution from the high momentum component
of the spectral function than the normalization integral 
$\int S({\bf k},E)d^3k dE$ since 
\begin{equation}
%\int S({\bf k_t},k_z=0,E)d^2{ k_t} dE
S(k_3=0)={1\over 2}\int S({\bf k},E){d^3k\over k} dE,
\end{equation}
leading to a strong enhancement of the small $k $ region.
This, in turn, implies that for
a given kinematics the contribution of the large excitation
energies ($\Delta E\ge 80-100 MeV$),
 which is predominantly due to the short range correlations,
is also insignificant.
Therefore,  we can use  mean field models to calculate the value of
$F(y=0)$ as measured in the $(e,e') $ processes for $Q^2\sim 1\, GeV^2$ where
inelastic contributions are
 still very small. Note that the account of the inelastic 
contributions allows for a good description
of the  $Q^2$ dependence of the 
ratio ${\sigma_{eA}(x,Q^2)/ \sigma_{eD}(x,Q^2)}$ at $x=1$,
over  a wide range of $Q^2$ \cite{fsscor}. In our calculations we 
used the 
Hartree-Fock-Skyrme model, which describes well many global properties
of nuclei e.g. the energy binding, the spectra of the single particle states, root mean square radii 
and the shape of the proton and neutron  matter distributions \cite{HFS}.
In this model, the spectral function is given by
\begin{equation}
S_{HF}({\bf k},E)=\sum_{\alpha} n_{\alpha}\delta (E-E_{\alpha})
{\vert \varphi_{\alpha}({\bf k}\vert}^2.
\end{equation}
The occupation probabilities for the filled nuclear levels are $n_{\alpha}=1$.
The integral which determines  cross section of the quasielastic 
scattering at $x\approx 1$ and $y=0$ is
\begin{equation}
S_{HF}(k_3=0)=
2\pi \int \limits_{0}^{\infty}\sum_{\alpha} n_{\alpha}
{\vert \varphi_{\alpha}(k)\vert}^2\,k\,d\,k.
\label{kper}
\end{equation}

\begin{figure}
\centering
\epsfig{file=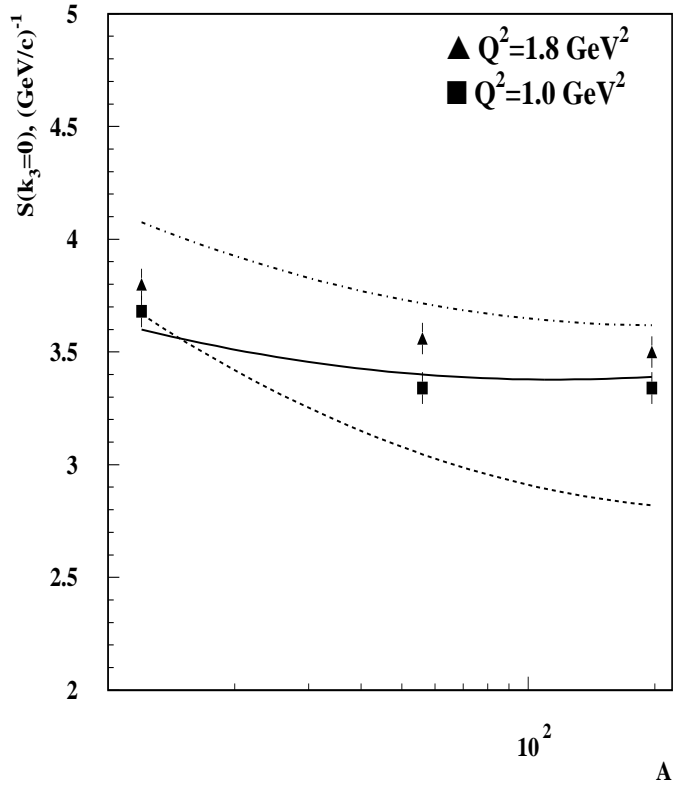,height=12cm,width=10cm}
\caption{Comparison of the scaling function $F_{A(e,e')X}(y=0,Q^2)$  
extracted 
from the data\cite{arrington} and
corrected for the inelastic contribution 
with the values of the integral (\ref{kper}) 
$S(k_3=0)$  calculated in the
Hartree-Fock-Skyrme model(solid line) and in the shell model used in
\cite{ne18,tjnaf,dutta}(dash-dotted line). The value of the integral (\ref{kper}) used
in \cite{ne18,tjnaf,dutta} 
in the denominator of Eq.(~\ref{tne18}) for extracting the nuclear transparency from the data
is shown by the dashed line. 
}
\label{inclusive} 
\end{figure}

The results of the calculation are presented  as
 the  solid curve in Fig.~\ref{inclusive}. 
They are  compared 
to the values of $F(y=0)$ extracted  
from the
inclusive $(e,e')$ 
 data of  \cite{arrington} in the
vicinity of $x=1$ at $y=0$. These experimental 
values were
corrected for a small contribution of   inelastic processes
using the analysis of \cite{fsscor}, which described well the onset 
of the dominance of the inelastic contribution with increase of 
$Q^2$. The correction is about 3\% (6\%)  for $Q^2=1.0(2.0)\, GeV^2$.
One can see that the data are described  very well without any adjusted
 parameters. 
Note that the IPSM spectral function which was used in the experimental
analysis of NE18 and E91-013 \cite{dutta1}
reproduces reasonably (dot-dashed line in Fig.~\ref{inclusive})
  the weak $A$-dependence of the data though it somewhat overestimates the absolute
value of the integral (\ref{kper}). At the same time 
%introducing 
an additional
 renormalization factor $f(A)$, Eq.(~\ref{corcor}), introduced in  \cite{ne18,tjnaf,dutta},  does a good
job for $A=12$ but leads to much stronger A-dependence of the
 integral(dashed line in Fig.~\ref{inclusive}) than suggested
 by our calculations 
and  by  the data.
We have also checked that the estimate of the integral (\ref{kper}) 
with the phenomenological  models of the
spectral functions, like, for example, those of \cite{CiofidegliAtti:1996qe},
which includes the 15\% contribution of the short-range correlations from 
region $k\ge 300\,MeV/c$ in the normalization integral for the spectral function,
coincides with our  results for (\ref{kper})  within 5\%. 
Also, in this model only 5\% of the integral originates from the 
region of $k \ge 300 MeV/c$.
This confirms our
conclusion about  the  
strong suppression of the short range correlation contribution
in the transverse kinematics of \cite{ne18,tjnaf,dutta} due to the
specific properties of the integral (\ref{kper}).
It is worth noting that already this comparison
gives a new  confirmation of  our result for the value of
the  quenching
factor $\eta\approx 0.9$, 
extracted in Ref. \cite{lsfsz} 
 from the comparison of the calculated 
momentum distribution in the C(e,e'p) process to that measured  
in the NE18 experiment at $Q^2=1\,GeV^2$.  

Hence we conclude that the wave functions we use are sufficiently realistic
to estimate the integral quantity entering the denominator of (\ref{tne18}) 
in the calculation of the transparency in the $A(e,e'p)$ reactions.

\section{Exclusive $(e,e'p)$ cross section}
\label{excl}

\begin{figure}
\centering
\epsfig{file=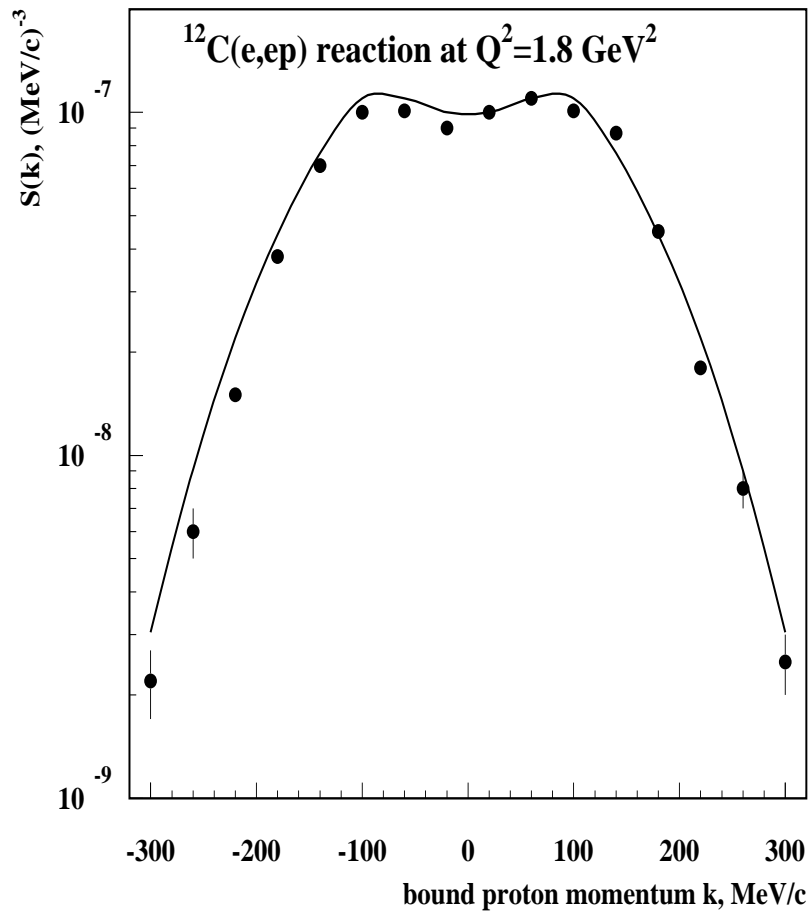,height=14cm,width=12cm}
\caption{Comparison of the calculated momentum distributions  for $^{12}C(e,e'p)$ 
at $Q^2=1.8\,GeV^2$ with E91-013
\cite{dutta} data.}
\label{cardis} 
\end{figure}

\begin{figure}
\centering
\epsfig{file=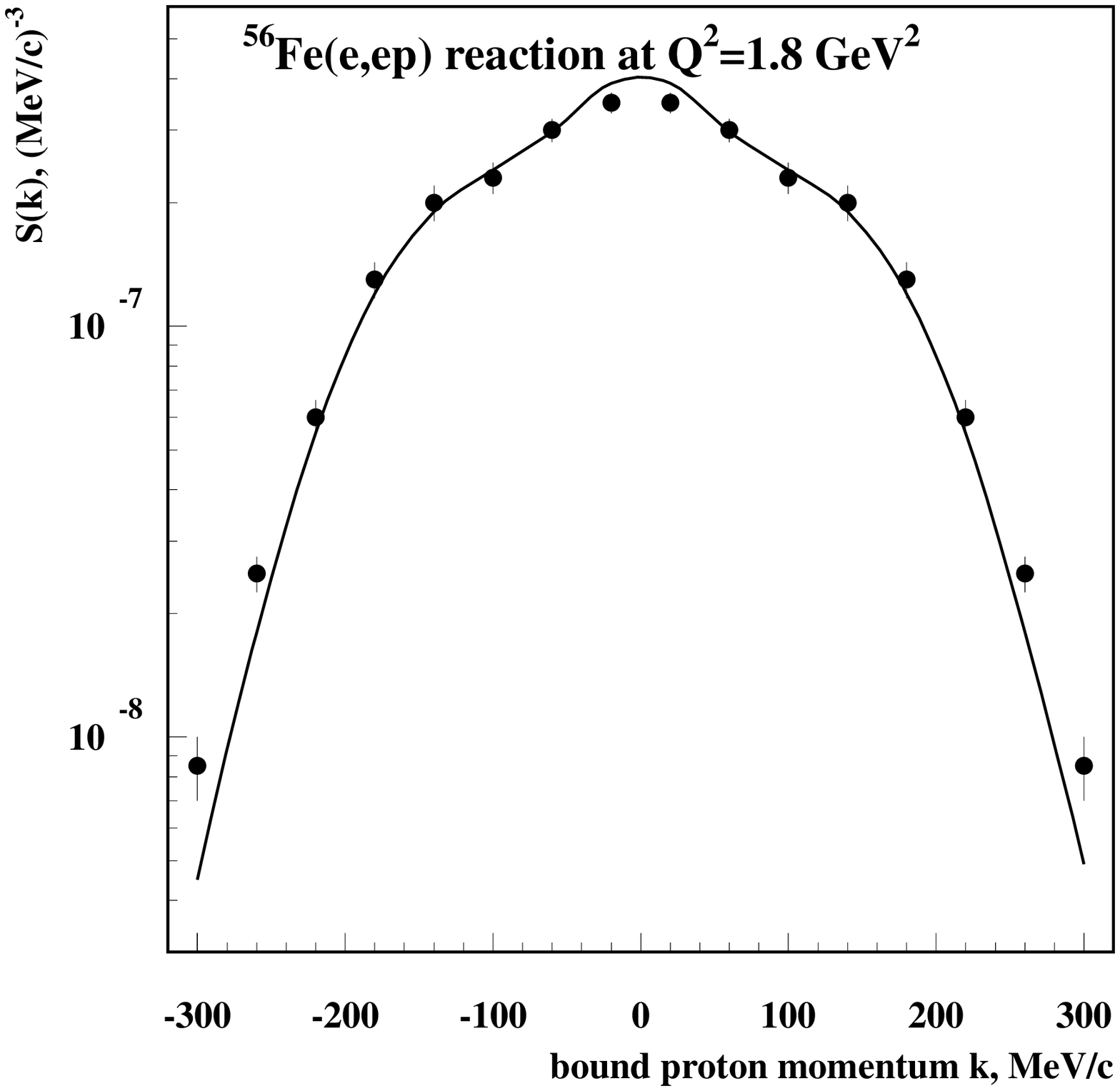,height=14cm,width=12cm}
\caption{Comparison of the calculated momentum distributions for $^{56}Fe(e,e'p)$
at $Q^2=1.8\,GeV^2$ with E91-013
\cite{dutta} data.}
\label{ferdis} 
\end{figure}

\begin{figure}
\centering
\epsfig{file=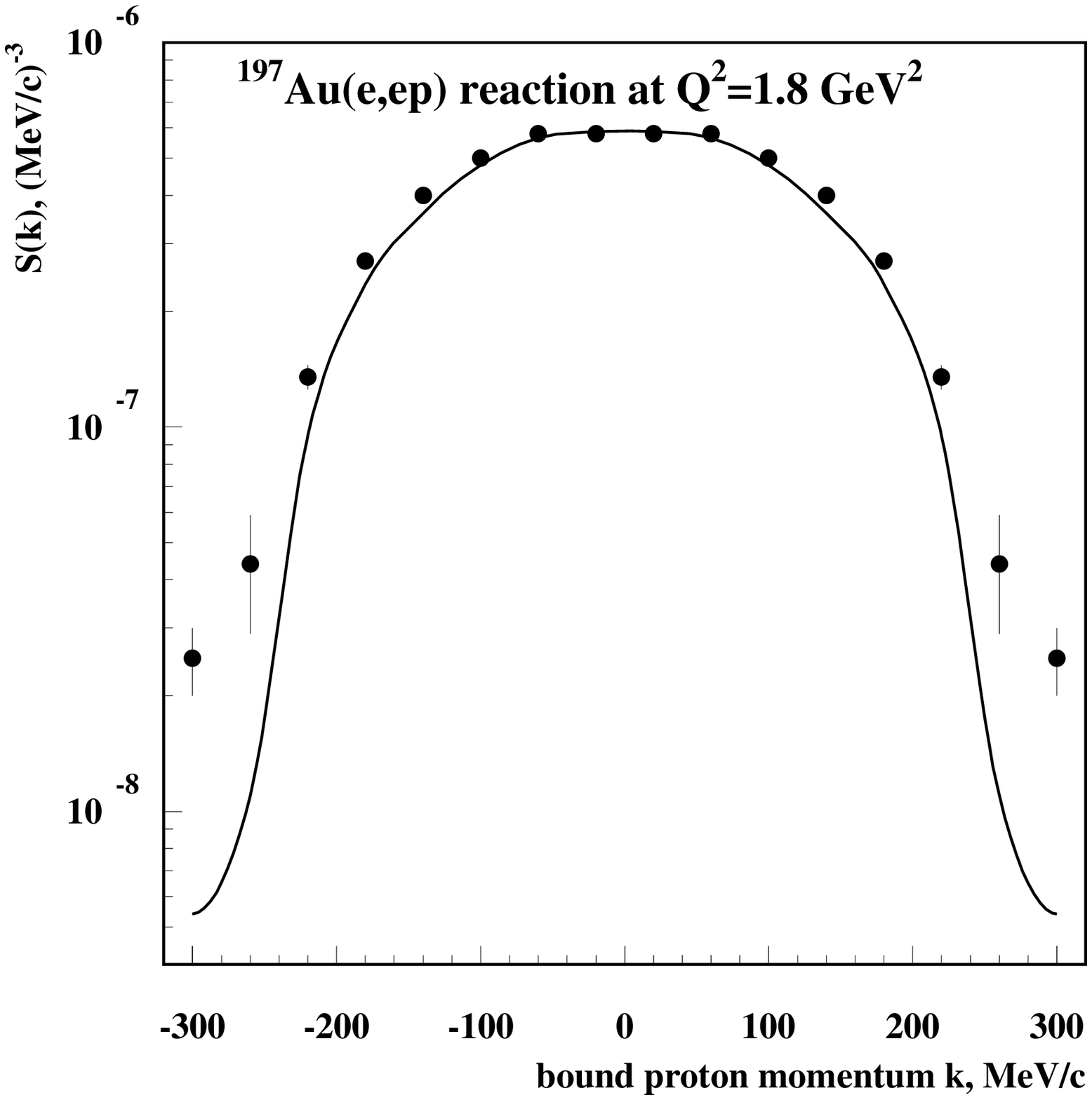,height=14cm,width=12cm}
\caption{Comparison of the calculated momentum distributions for $^{197}Au(e,e'p)$
at $Q^2=1.8\,GeV^2$ with E91-013
\cite{dutta} data.}
\label{goldis} 
\end{figure}

   A  more stringent test of the wave functions and 
interpretation of the data 
can be reached using differential data from E91-013 \cite{dutta}. 
The differential cross section of the $(e,e'p)$ were calculated
using the Hartree-Fock-Skyrme wave functions and the Glauber type model of the FSI for
$(e,e'p)$ reactions\cite{frank1,frank2} \footnote{Note that we neglect in this 
calculation corrections due to effects of the short-range correlations
in the propagation of the knocked out nucleon through the nucleus, which
are known to be a small $\leq few \%$ correction for small struck
nucleon momenta\cite{Frankfurt:1995rq,Nikolaev:1993sj}.}.
The cross section can be represented 
in the same form as the impulse approximation with a
substitution of the spectral function by the sum of 
 the distorted shell momentum distributions 
given by 
\begin{eqnarray}
 S^{fsi}_{HF}({\bf k})=\sum_{\alpha}n_{\alpha}
 \biggl \vert
 \int \,d\,{\bf r}\,
 e^{[i{\bf k\cdot r}]} \, \phi_{\alpha}({\bf r})  
 \biggl \langle (A-1)\biggl \vert
 \vert 
 \prod_{j=1}^{A-1}[1- 
 \Gamma({\bf b-b_j})\theta(z-z_j)
 ]\biggr \vert
 (A-1)\biggr \rangle 
 \biggr \vert ^2.
\label{eq:cseep}
\end{eqnarray}
Here $$\Gamma({\bf b}-{\bf b_j})=(2\pi ip)^{-1}\int d^2q_t \, 
exp[-{\bf q_t}\cdot ({\bf {b-b_j}})]\cdot f_{NN}({\bf q_t}),$$
and  the $NN$ amplitude for  high energy protons is given by 
the expression
$$f_{NN}({\bf q_t})=\frac {p\sigma^{tot}_{pN}
(1-i\kappa_{NN})}{4\pi}e^{-\frac {B^2q^2_t}{2}}.$$
The values of the total proton-nucleon cross section $\sigma^{tot}_{pN}$,
the slope parameter $B$ 
and the real-to-imaginary ratio  $\kappa_{pN}$  of the pN amplitude are well
known\cite{vorobiov}
 for the 970 MeV protons corresponding to the $(e,e'p)$ reaction
 at $Q^2=1.82\,GeV^2$.
 Results of our calculation for the distorted momentum distributions
using eq.\ref{eq:cseep} and the HF spectral function
are compared to the data of 
\cite{dutta}
in Figs.~\ref{cardis},~\ref{ferdis},~\ref{goldis}. 
Taking into account that our calculations do not comprise 
any free parameters,
one  observes a   fair agreement with experimental
data, at least for    momenta of the
bound proton $\le$ 200 MeV/c. 
A discrepancy at momenta above 200 MeV/c, which increases with A, can be
considered  as an evidence for the elastic incoherent  rescattering processes
for the outgoing nucleon (this effect will be considered elsewhere).

\section{Inclusive transparency in $A(e,e'p)$ reaction.}
\label{incl}

The E91-013 experiment at Jlab\cite{tjnaf}   which we discussed in the previous section
also reported new values for the nuclear transparency which are 
  consistent with the NE18 data  but has
somewhat better accuracy.
In the previous sections we have demonstrated
that both the numerator and the denominator in the definition
 of the transparency in Eq.(\ref{tne18})
in the kinematics of the NE18 and E91-013
are  sensitive to assumptions about the spectral function, but 
are  strongly constrained by 
 the $(e,e')$ and $(e,e'p)$ data at $x\sim 1 $ and $Q^2\sim (1\div 2) \,GeV^2$.
This allows us to treat ${\bf T} $ with much smaller uncertainties than before
and use it to obtain a suplementary
evidence of the single particle strength restoration
at high  momentum transfer.

\begin{figure}
\centering
\epsfig{figure=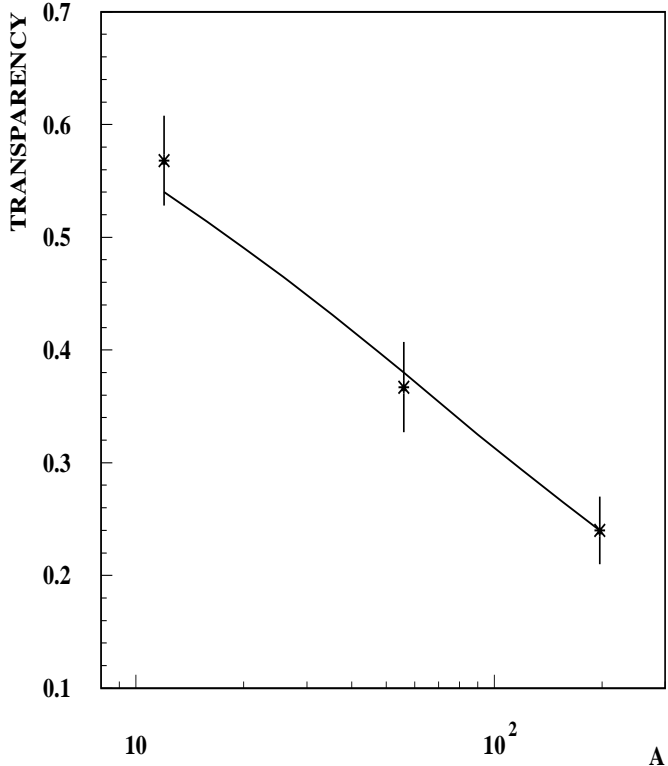,height=12cm,width=10cm}
\caption{Comparison of the transparency calculated in the Glauber model for 
the FSI (solid line) with the measured nuclear transparency\cite{tjnaf}
corrected for the difference in the impulse approximation cross section.} 
%in accordance with Eq.(~\ref{renorm})}.
\label{transp}
\end{figure}

We obtain theoretical value of ${\bf T}$ from Eq.~\ref{tne18}
by using the spectral functions $S_HF(k_3=0)$ (Eq.~\ref{kper}) and  
by integrating Eq.(\ref{eq:cseep})  to account for the FSI 
in the kinematics of \cite{dutta}. The result of calculation is
 presented by solid curve in Fig.\ref{transp}.
To compare these results with the data in a way consistent with our finding in 
section \ref{norma} we need to use the impulse approximation
cross section consistent with the results presented 
in Fig.~\ref{inclusive} and with the theoretical calculation of ${\bf T}$. 
The simplest procedure is 
to correct the transparency values
presented in \cite{ne18,tjnaf,dutta}
using in the denominator of Eq.(\ref{tne18}) the value 
of the cross section in the impulse approximation 
given by our spectral function which differ from the value used in \cite{ne18,tjnaf,dutta}
by the factor ${S_{IPSM}(k_3=0)f(A)/ S_{HF}(k_3=0)}$
(the factors are 1.02 for carbon, 0.896 for iron and 0.83 for gold).
This leads to the corrected experimental values of the transparency
shown  in Fig.\ref{transp} which are in a good agreement with results of calculations.
We want to emphasize that this modification of the values of ${\bf T}$,
presented in \cite{ne18,tjnaf,dutta}, arises solely due to the change of 
the theoretical estimate of the impulse approximation cross section used 
in \cite{ne18,tjnaf,dutta} to ensure consistency with results of the (e,e') measurements.

%We see that there is a good agreement between the calculation and the corrected data.

\section{The $Q^2$ dependence of quenching}

At first  glance, the comparison with the data performed in
previous Sections
 leaves no room for the presence 
of  single particle strength quenching at $Q^2\ge 2 GeV^2$.
However to make the final conclusion one should carefully  take into
account  experimental errors and  uncertainties  of the 
calculations. Generally, the  accuracy of the Glauber approach in
the description of the  proton-nucleus interaction
in high energy kinematics of the (e,e'p) reaction  
at $Q^2 \geq 2\, GeV^2$ is estimated to be small (a  few \%)   as 
long as no new physics like color transparency is present.
There exists also a few \%  uncertainty 
due to the use in the calculation of 
a definite set of the Hartree-Fock wave functions
and neglecting effects of the short-range correlations
in the calculation of the normalization integral (\ref{kper}).
Hence, a possibility of the  quenching  of 
order $\approx 10\%$ cannot be excluded.
However, this is definitely   much smaller than
  $\eta(^{12}C)= .57\pm .02$
determined from the low $Q^2$ data \cite{lsfsz}.

It should be noted  that the analysis of (e,e'p) data is
more definitive  at high energy and high  momentum transfer than in 
the low energy
kinematics. 
The kinematical off-shell effect in the  $ep$ vertex
due to the Fermi motion of nucleon,  studied e.g. 
by De Forest\cite{deforest},
is minimized in the high energy limit. Also, the
 renormalization of the
$ep$ vertex  due to the inability of a  low $Q^2$ photon to  resolve
the  short-range  and 
 long-range  correlations of interacting proton with the rest of the  nucleons
 is evidently more essential in the low $Q^2$ 
 kinematics. Within the quasiparticle approach,
such a renormalization can be performed  by using the
form factor of a quasiparticle which is
softer than for a free nucleon, because 
at low  resolution a
 low momentum bound nucleon in the nuclear 
medium is dressed by a cloud of virtual nuclear excitations.    
With increase of  the momentum transfer  above 
the Fermi-momentum of the
bound nucleon, $k_F\approx (220\div 260) \,MeV/c$, this 
renormalization of the electron-proton vertex disappears,
 and we deal 
with the form factor of a free nucleon.

Besides, taking into  account  the FSI at low energies is more 
cumbersome because
one needs to deal with 
the optical potentials which are determined from 
 fits to  the proton-nucleus elastic scattering data.
Such a treatment ignores the  
difference  in  the space geometry of  proton-nucleus elastic scattering,
dominated by the interaction with the nuclear surface,
and the proton propagation
from the nucleus interior in the electron induced nucleon knockout reaction.

To summarize, we have demonstrated,
based on the  joint analysis of the exclusive A(e,e'p) and A(e,e')X data
at $Q^2\geq 1\,GeV^2$,  
that  the actual quenching factor which enters into cross sections
of the exclusive quasielastic processes differs from the one  used in 
\cite{ne18,tjnaf, dutta} and
is practically insensitive to the probability of the short-range 
nucleon correlations in nuclei.  
We found further  evidence for the dependence of the single particle
strength quenching in the exclusive (e,e'p) reactions on the momentum transfer.
The strong effect (about 40 $\%$), observed in the low energy phenomena, 
practically disappears with increase of $Q^2$, when a probe resolves the quasiparticle
structure of the nucleon  due to
the long-range correlations
inside the nuclear medium. 
 In the discussed
transverse kinematics only a very modest
quenching (less than 10$\%$) may survive
 in the exclusive (e,e'p) reaction
at high $Q^2$ and $|k|\le 300 MeV/c$.

 A strong $Q^2$ dependence of quenching comes very naturally in the
Fermi liquid theory \cite{LL,Migdal,Brown}  and really represents  
the  generic property of 
fermionic systems where the interaction between fermions is described by a 
renormalizable theory\cite{Wilson}, like QED or QCD,
 since in this case the wave functions
 of constituents depend strongly on the resolution scale.

A high precision measurements of  A(e,e') scattering 
and the differential cross sections of the exclusive A(e,e'p) reactions
at $Q^2$ in the range $(1\div 2)\,GeV^2$
would be very useful for an accurate estimate of the 
quenching effect,  and to determine the  experimental values of
nuclear 
transparency.  
 The kinematics $\Delta E \le 80 MeV, |k|\le 300 MeV/c$ appear to
be optimal for searches 
of   color transparency at moderate and high momentum 
transfers in the $(e,e'p)$ reactions, in order
to understand the phenomenon of expanding of  small
size quark configurations in hard processes.

We thank R.Ent for stimulating questions and comments,
D.Dutta for the information about the spectral function used in
the E91-013 experiment,
 M.Sargian  for discussion
of inelastic contributions. We also thank  B.Birbrair, L. Lapik\'{a}s and  G. van der Steenhoven
for numerous discussions of the relation between low $Q^2$ and high $Q^2$
(e,e'p) processes. Research of M.S. and M.Z. was supported in part
by the U.S. Department of Energy, research of L.F. was supported by
the Israeli Academy of Science.


\begin{thebibliography}{98}





\bibitem{mogad}
S.~Frullani and J.~Mougey,
%``Single Particle Properties Of Nuclei Through (E, E' P) Reactions,''
Adv.\ Nucl.\ Phys.\  {\bf 14}, 1 (1984).



\bibitem{SWH} G.~ van~ der~ Steenhoven and P.~K.~A.~ de~ Witt~ Huberts,
in Modern Topics in Electron Scattering,
Eds. B. Frois and I. Sick, World Scientific, 1991, p. 510



\bibitem{Kelly:1996hd}
J.~J.~Kelly,
%``Nucleon knockout by intermediate-energy electrons,''
Adv.\ Nucl.\ Phys.\  {\bf 23}, 75 (1996).


\bibitem{Migdal} 
A.~B.~Migdal. Theory of Finite Fermi System
and Application to Atomic Nuclei (Interscience, New York, 1967).


\bibitem{Dickhoff:1992um}
W.~H.~Dickhoff and H.~Muther,
%``Nucleon properties in the nuclear medium,''
Rept.\ Prog.\ Phys.\  {\bf 55}, 1947 (1992).



\bibitem{lsfsz}
L. ~Lapikas, G.~van ~der~ Steenhoven, L.~Frankfurt et.al.
Phys.\ Rev\ C{\bf 61},064325, (2000).


\bibitem{ne18}
N.C.R Makins \etal,
Phys.\ Rev.\ Lett.\ {\bf 72},  (1986), (1994);\\
T.~G.~ O'Neill \etal,
Phys.\ Lett.\ {\bf B351},  (87), (1995);\\
T.G. O'Neill,
PhD Thesis, California Institute of Technology (1994) unpublished,
and N.~C.~R. ~Makins,
PhD Thesis, Massachusetts Institute of Technology (1994) unpublished



\bibitem{tjnaf}
D.~Abbott \etal
Phys.\ Rev.\ Lett. {\bf 80}, 5072, (1998)





\bibitem{CT}
A. ~H. ~Mueller, in:
{\it Proc. of the XVII Rencontre de Moriond, 1982},
ed. J. Tran Thanh Van
(Editions Fronti\`{e}res, Gif-sur-Yvette, France, 1982) p. 13.



\bibitem{Brodsky}
S. ~J.~ Brodsky, in:
{\it Proc. of the Thirteenth International Symposium on
Multiparticle Dynamics}, eds. W. Kittel, W. Metzger, and A.
Stergiou (World Scientific, Singapore, 1982) p. 963.


\bibitem{Frankfurt:1994hf}
L.~L.~Frankfurt, G.~A.~Miller and M.~Strikman,
%``The Geometrical color optics of coherent high-energy processes,''
Ann.\ Rev.\ Nucl.\ Part.\ Sci.\  {\bf 44}, 501 (1994)
[hep-ph/9407274].


\bibitem{frank1}
L. ~Frankfurt, M.~Strikman, M.~Zhalov
Nucl.\ Phys.\ {\bf A515}, 599, (1990).

\bibitem{frank2}
L. ~Frankfurt, M.~Strikman, M.~Zhalov, Phys.\ Rev.\ {\bf C50},2189,(1994).



\bibitem{deforest}
T. ~de ~Forest
Nucl.\ Phys.\ {\bf A392},  232, (1983).


\bibitem{fsscor}
L. ~Frankfurt, M. ~Strikman, D. ~Day, M. ~Sargsyan
Phys.\ Rev.\  C{\bf 48}, 2451, (1993). 



\bibitem{vorobiov}
G.~D.~Alkhazov \etal, Phys.\ Rep.\ {\bf 42},89,(1978).



\bibitem{arrington}
J.~Arrington {\it et al.},
%``Inclusive electron nucleus scattering at large momentum transfer,''
Phys.\ Rev.\ Lett.\  {\bf 82}, 2056 (1999)
[nucl-ex/9811008];
J.~R.~Arrington, PhD Thesis, Caltech, 1998, unpublished.






\bibitem{dutta}
D.~ Dutta, PhD Thesis, Northwestern University (1999), UMI-99-32159-mc


\bibitem{Golubeva:1998at}
Y.~S.~Golubeva, L.~A.~Kondratyuk, A.~Bianconi, S.~Boffi and M.~Radici,
%``Nuclear transparency in quasielastic A(e,e' p): Intranuclear 
%cascade  versus eikonal approximation,''
Phys.\ Rev.\  {\bf C57}, 2618 (1998)
[nucl-th/9712040].







\bibitem{HFS}
D.~Vautherin and D.~M.~Brink,
%``Hartree-Fock Calculations With Skyrme's Interaction. 1. Spherical Nuclei,''
Phys.\ Rev.\  {\bf C5}, 626 (1972);\\
M.~Beiner, H.~Flocard, N.~Van Giai, and P.~Quentin,
Nucl.\ Phys.\ {\bf A238}, 29, (1975).


\bibitem{dutta1} D.Dutta, Private communication.


\bibitem{CiofidegliAtti:1996qe}
C.~Ciofi degli Atti and S.~Simula,
%``Realistic model of the nucleon spectral function in few and many nucleon systems,''
Phys.\ Rev.\  {\bf C53}, 1689 (1996)
[nucl-th/9507024].











\bibitem{Frankfurt:1995rq}
L.~L.~Frankfurt, E.~J.~Moniz, M.~M.~Sargsian and M.~I.~Strikman,
%``Correlation effects in nuclear transparency,''
Phys.\ Rev.\  {\bf C51}, 3435, (1995)
[nucl-th/9501019].



\bibitem{Nikolaev:1993sj}
N.~N.~Nikolaev, A.~Szczurek, J.~Speth, J.~Wambach, B.~G.~Zakharov and V.~R.~Zoller,
%``Correlation effects in the final state interaction for quasielastic (e, e-prime p) scattering,''
Phys.\ Lett.\  {\bf B317}, 281 (1993).







\bibitem{LL} 
L.~D.~ Landau and E.~M.~ Lifshitz, Statistical Physics,part 1\\
(Course of Theoretical Physics, v. 5) Pergamon Press, 1980.








\bibitem{Brown}
G.~E.~ Brown,
   Unified Theory of Nuclear Models and Forces.  North-Holland, 316p., 1971





\bibitem{Wilson}
K.~G.~Wilson,
 Phys. Rev.  {\bf B4} (3174), 1971. 





\end{thebibliography}
\end{document}